\documentclass[graybox,natbib,nosecnum]{svmult}
\bibpunct{(}{)}{;}{a}{}{,} 

\pdfoutput=1   

\usepackage{helvet}         
\usepackage{courier}        

\usepackage{makeidx}         
\usepackage{graphicx}        
\usepackage{multicol}        
\usepackage[bottom]{footmisc}
\usepackage[normalem]{ulem}	
\usepackage{hyperref}  

\usepackage{soul}   

\makeatletter
\providecommand*{\toclevel@title}{0} 
\providecommand*{\toclevel@author}{0} 
\makeatother

\newcommand*\aap{A\&A}

\newcommand*\aj{AJ}

\newcommand*\apj{ApJ}
\newcommand*\apjl{ApJ}

\newcommand*\apjs{ApJS}

\newcommand*\araa{ARA\&A}
\newcommand*\azh{AZh}

\newcommand*\icarus{Icarus}

\newcommand*\jgr{J Geophys Res}

\newcommand*\mnras{MNRAS}

\newcommand*\nat{Nature}

\newcommand*\procspie{Proc SPIE}

\newcommand{\hbindex}[1]{{#1}}  

\makeindex             

\begin{document}

\title*{Exoplanets and SETI}
\author{Jason T.\ Wright}
\institute{Jason T.\ Wright \at Department of Astronomy \& Astrophysics and
Center for Exoplanets and Habitable Worlds \\525 Davey Laboratory, 
The Pennsylvania State University, 
University Park, PA, 16802, USA\\ 
Visiting Associate Professor, Breakthrough Listen Laboratory, Department of Astronomy, University of California, Berkeley, CA, 94720, USA 
\\ \email{astrowright@gmail.com}}

\maketitle

\abstract{ 
The discovery of exoplanets has both focused and expanded the search for extraterrestrial intelligence. The consideration of Earth as an exoplanet, the knowledge of the orbital parameters of individual exoplanets, and our new understanding of the prevalence of exoplanets throughout the galaxy have all altered the search strategies of communication SETI efforts, by inspiring new ``Schelling points'' (i.e.\ optimal search strategies for beacons). Future efforts to characterize individual planets photometrically and spectroscopically, with imaging and via transit, will also allow for searches for a variety of technosignatures on their surfaces, in their atmospheres, and in orbit around them. In the near-term, searches for new planetary systems might even turn up free-floating megastructures.
}

\section{Introduction }
The discovery and characterization of exoplanets is central to astrobiology: exoplanets are the most natural locations to search for life elsewhere in the universe. One approach is to move toward the detection of {\it biosignatures} that might be produced extraterrestrial life; the search for extraterrestrial intelligence (SETI) focuses instead of {\it \hbindex{technosignatures}} that might be produced by {\it intelligent} life.

Many proposed technosignatures of extraterrestrial civilizations in addition to  electromagnetic communications might be observable today or in the foreseeable future, including city lights, atmospheric pollutants, waste heat, and the transits of megastructures. The search for such technosignatures is often called \hbindex{\it artifact SETI} (distinguished from \hbindex{\it communication SETI}).

Indeed, these civilizations need not be active today to be detectable.  \citet{Freeman75} and \citet{Campbell06} proposed artifact SETI as a form of interstellar ``archeology,'' suggesting that we might find the remnants of {\it extinct} extraterrestrial civilizations, a theme extended by \citet{Carrigan12}, and \citet{Stevens16}.

\section{Exoplanets as Schelling Points in Communication SETI}

Two of the many dimensions of the vast parameter space of communication SETI \citep[e.g.][]{Tarter01} are {\it when} to observe (or transmit to) a given target, and {\it which directions} to target at a given moment. If one assumes that the search for and transmission of of deliberate signals \citep[``beacons,''][]{Dixon73} is a mutual endeavor, then one can turn to game theory's analysis of the problem of a cooperative game in which the players cannot communicate.

\citet{schelling1960strategy} described \hbindex{\it focal points} (better called \hbindex{\it Schelling points} in astronomy to avoid ambiguity) as mutually obvious locations in the strategy space of such a game. His examples involved finding a person in a city who is also looking for you, and radio SETI, citing \citet{SETI}. Guessing the times and places to meet in the city, and guessing the frequencies to tune to in radio SETI, are superior strategies to random ones. In the city, this might include the locations of famous landmarks and times that bells chime or other coordinated actions occur; in radio SETI this might mean astrophysically significant frequencies and their multiples. \citet{Makovetskii80} called this a ``mutual strategy of search'' for ``synchrosignals''\citep{Makovetskii77}, and \citet{Filippova91} described this concept as a ``convergent strateg[y] of mutual searches'' for SETI (both apparently unaware of \citeauthor{schelling1960strategy}'s prior art).

\subsection{Where to Observe}

Exoplanets form a natural Schelling point, and since communication SETI efforts typically spend more effort on targets where life is more likely to be found, they make natural communication SETI targets. Since the beginning of the field, communication efforts have focused on Sun-like stars likely to host habitable planets \citep[some recent examples of such target lists include][]{Henry95,Turnbull03a,Turnbull03b}.

The prospect of alien civilizations detecting Solar System planets as exoplanets inspired similar thoughts. \citet{Filippova88} called the ecliptic an ``attractor for SETI'' because Earth would appear to transit the Sun from stars there. \citet{Corbet03} argued that {\it all} stars at opposition (i.e.\ those seeing Earth at inferior conjunction, not just those  seeing Earth transit) should be searched for this reason.

As \citet{Bracewell79} predicted, the discovery of individual exoplanets, especially rocky planets and those in the Habitable Zones of their host stars \citep{Kasting93}, has naturally focused efforts on them and their orbits \citep{Siemion2013,Panov14,Harp16_exoplanets}.

That said, the discovery of many exoplanets has also shown that the occurrence rates of rocky exoplanets in the Habitable Zones of stars is so high \citep[of order 10\%, and likely higher][]{Traub12,Petigura13,Dressing15} that no stars should be neglected simply because they have not had any of their habitable planets discovered {\it yet}. This is why many surveys have returned to the original strategy of surveying stars independent of their known planet status \citep{Maire16,Isaacson17}.

\subsection{When to Observe}

As suggestion for a temporal Schelling point, \citet{Pace75} suggested observing binary stars during periastron and apastron. \citet{Makovetskii77} suggested sending and listening for signals coincident with other predictable astronomical phenomena, targeting those and opposites part of the sky. This transmission strategy would mean that even astronomers observing these phenomena for non-SETI purposes might detect the signal.  

Again, considerations of the Solar System objects as exoplanets has sharpened the discussion. \cite{Singer82} suggested using the times of maximum displacement of the Sun by Jupiter as Schelling points; \citet{Filippova91}, and later \citet{Shostak04} suggested that we search stars along the ecliptic during the time Earth would appear to transit from the transmitter's perspective, although this strategy requires either the transmitter or the receiver to make adjustments for light travel time, which requires precise knowledge of the distance between them.

The actual discovery of transiting exoplanets has allowed for an even more focused approach: searching for signals during the time the {\it exoplanet} transits. \citet{Kipping16} argued ``the time of transit provides a natural communication window, analogous to water hole in radio SETI \citep{Oliver79},'' (i.e.\ a Schelling point). This strategy has the  advantage that distances to the targets need not be known (the light travel time is the same for the signal and the light of the transit). By an extension analogous to that of \citet{Corbet03}, one might search any planet during its inferior conjunction with its star.

\section{Technosignatures on Exoplanets and the Host Stars}

If alien civilizations are not broadcasting beacons we are meant to find, then the Schelling point concept is irrelevant, and the questions of where and when to look revolve around different questions. For communication SETI, this means when and where are we most likely to intercept leaked emission. For instance, \citet{Siemion2013} proposed eavesdropping on planet-planet communications which is best performed when two inhabited planets in an edge-on multiplanet system are in conjunction and transmissions from the farther to the nearer planet will be inadvertently directed at Earth. \citet{Guillochon15} proposed looking for leaked energy from propulsion systems at the same time for that same reason.

On the artifact SETI side, although the direct imaging of large structures on exoplanetary surfaces would require angular resolutions too far in the future for even this work to consider, other options exist \citep{Kreidberg16,Cowan17}. \citet{Campbell06} and later \citet{Schneider10} proposed that the direct imaging of exoplanets presents special opportunities for the detection of technosignatures. Surface maps can be constructed using a planet's rotationally modulated brightness \citep{Kawahara10}. This is even true when they are not directly imaged, both in photometry \citep{Knutson07}, and from their secondary eclipse light curves \citep{Majeau12,deWit12}.

\runinhead{Waste Heat}

A mid-infrared map with sufficiently high sensitivity might allow one to conduct a \hbindex{waste heat} search for civilizations \citep{Dyson60,carrigan09a,GHAT2} by looking for industrial heat signatures on the planetary surface. For instance, \citet{kuhn2015global} suggested that a 70m telescope might be sufficient to detect the rotationally modulated localized output of industry on an Earth-like planet for a civilization with $\sim 25$ times the energy supply of humanity (which is equal to about 1\% of light the planet intercepts from its star).

\runinhead{Artificial Illumination}

\citet{Schneider10} suggested that artificial light sources might be detectable on the night sides of planets, and \citet{Loeb11} pointed out that proposed versions of space telescopes might be able to detect such ``city lights'' via direct imaging if they are a few times more powerful than those of Earth.

\citet{Kipping16} recommend searching for laser emission at the time of transit, especially in the form of anomalous transit light curves or transit spectra.  They suggest that a civilization might use such lasers to attract attention when we are studying their planet's transit, or might use it to ``cloak'' their planet's transit light curve or spectrum biosignatures.

\runinhead{Spectroscopic Detection of Pollution}

Exoplanetary atmospheres are amenable to spectroscopy in several ways, including in thermal emission and via reflection/absorption of starlight, and via transit spectroscopy.  These techniques can all probe different wavelengths and atmospheric depths, and so potentially probe a variety of atmospheric technosignatures.

\citet{Schneider10} suggested that technosignatures might be present in the atmosphere in the form of unnatural chemical substances, perhaps due to \hbindex{pollution}, such as our chlorofluorocarbons (CFCs) or photovoltaic arrays \citep{Lingam17}. \citet{Lin14} calculated that over 1 day of integration with the {\it James Webb Space Telescope} might be able to detect CFCs at only 10 times their current concentrations on Earth. \citet{Stevens16} also presents several scenarios that might only be just detectable and recognizable if we were to happen to catch a cataclysm like those we fear for Earth at the moment it happened, cosmically speaking. For instance, they argue that the signatures of global nuclear war, including gamma rays, the chemical effects of radioisotopes and the heat of nuclear weapons, and the following ``nuclear winter'' might all be detectable with sufficient precision of imaging and spectroscopy across the EM spectrum.

More likely, perhaps, than alien civilizations producing the same sorts of pollution that humans do or might create in the near future, would be the creation of clearly artificial chemicals of utilities that are unclear to us. An unrecognizable spectral signature might pique interest for further study, as astronomers travel down the long road of exclusion of natural causes \citep{GHAT1}.

Not only the planet might host pollution. Despite the folly inherent in  suggestions to launch humanity's waste into space, advanced civilizations might use their {\it star} as a dumping ground for dangerous or otherwise unwanted substances. \citet{Whitmire80} suggest it might be done as a way to dispose of fissile waste, and \citet{Stevens16} suggests such dumping might even result in a detectable environmental catastrophe. On the other hand, \citet{shklovskii1966} note independent suggestions by Drake and Shklovski{\u\i} that such pollution might be created deliberately as a ``beacon.''

Regardless of the reason for its presence, in most stars such pollution would be atomized and ionized by the star's envelope, and so would be only detectable via abundance anomalies, especially for elements or isotopes that are inherently rare in stars.  \citet{Whitmire80} suggest praseodymium as a good tracer of artificial nuclear reactions. Przybylski's Star,\footnote{Roughly pronounced (p)shi-BILL-skee, with a weak initial "p" as in the interjection ``pshaw''} \citep{Przybylski61,Cowley00} which shows evidence of high concentrations of numerous lanthanides and short-lived actinides in its atmosphere, is occasionally mentioned as a SETI candidate under this category (although never, that I can find, in the refereed literature.) 

\section{Megastructures}

\citet{Dyson60} suggested that advanced alien civilizations might intercept large amounts of starlight to power themselves, and be detectable by their waste heat in the mid-infrared. Dyson had in mind that the total infrared flux from the star would be anomalously large, but future direct imaging efforts may have the sensitivity to detect planet-sized starlight-blocking structures in reflected light or thermal radiation directly. 

\citet{Arnold05} applied this reasoning to the {\it Kepler} space observatory, noting that its photometric precision was sufficient to distinguish planets from planet-sized objects with non-circular aspect ratios. \citet{Arnold05} further noted that such structures might serve not just as power collectors but as highly efficient beacons (efficient in terms of the ergs per bits required to transmit information, since they passively {\it block} EM radiation being emitted anyway by the star).

Other artifacts that might be discovered include large satellites of inhabited planets \citep{Korpela15}, very large shields \citep{Forgan13}, or rings from a cataclysm \citep{Stevens16} such as a runaway collisional cascade of artificial satellites 
\citep[``Kessler syndrome''][]{Kessler78} or even total planetary destruction.

\citet{GHAT4} enumerated ten ways that planet-sized artificial structures (``\hbindex{megastructures}'') might be distinguished in a transiting planet survey from planets including anomalous light curve shapes, colors, transit timings, and follow-up signals.  They also noted natural confounders in each category; indeed each of their ten signatures is already being sought (and found) as a way to measure planetary masses (e.g.\ via transit timing variations), planetary clouds, exomoons, exorings, stellar and planetary oblateness, stellar limb and gravity darkening, atmospheric escape, starspots, orbital eccentricity, and circumstellar disks.  

The list of the confounders for these ten signatures are actually a good list of the most exciting topics of exoplanetary research in the future. Artifact SETI can thus ``piggyback'' on work likely to happen in the future, anyway, as natural anomalies are discovered in the course of exoplanetary science. Indeed, the pulsar planets \citep{Wolszczan92} show that we can expect to find planets, and thus, potentially, life (indigenous or not), around all types of stars. The search for megastructures should thus include pulsars \citep{Osmanov16}, X-ray binaries \citep{Imara17}, and other systems.

\begin{acknowledgement}
I think Jean Schneider for soliciting this document, Scott Colby for helpful conversations about the game theory of SETI and Schelling points, and David Kipping. I thank Jill Tarter, Seth Shostak, Manasavi Lingam, Andrew Howard, and Geoff Marcy for their assistance tracking down references to some ideas. I thank the Berkeley SETI Research Center for their support and encouragement during my sabbatical, and Breakthrough Listen, part of the Breakthrough Initiatives sponsored by the Breakthrough Prize Foundation,\footnote{\url{http://www.breakthroughinitiatives.org}}, for partially supporting this work. 

\end{acknowledgement}

\bibliographystyle{spbasicHBexo}  

\begin{thebibliography}{59}
\providecommand{\natexlab}[1]{#1}
\providecommand{\url}[1]{{#1}}
\providecommand{\urlprefix}{URL }
\expandafter\ifx\csname urlstyle\endcsname\relax
  \providecommand{\doi}[1]{DOI~\discretionary{}{}{}#1}\else
  \providecommand{\doi}{DOI~\discretionary{}{}{}\begingroup
  \urlstyle{rm}\Url}\fi
\providecommand{\eprint}[2][]{\url{#2}}

\bibitem[{{Arnold}(2005)}]{Arnold05}
{Arnold} LFA (2005) {Transit Light-Curve Signatures of Artificial Objects}.
  \apj 627:534--539

\bibitem[{{Bracewell} and {MacPhie}(1979)}]{Bracewell79}
{Bracewell} RN {MacPhie} RH (1979) {Searching for nonsolar planets}. \icarus
  38:136--147

\bibitem[{{Campbell}(2006)}]{Campbell06}
{Campbell} JB (2006) {Archaeology and direct imaging of exoplanets}. In: {Aime}
  C {Vakili} F (eds) IAU Colloq. 200: Direct Imaging of Exoplanets: Science \&
  Techniques, pp 247--250, \doi{10.1017/S1743921306009392}

\bibitem[{{Carrigan}(2009)}]{carrigan09a}
{Carrigan} RA Jr (2009) {IRAS-Based Whole-Sky Upper Limit on Dyson Spheres}.
  \apj 698:2075

\bibitem[{{Carrigan}(2012)}]{Carrigan12}
{Carrigan} RA Jr (2012) {Is interstellar archeology possible?} Acta
  Astronautica 78:121--126

\bibitem[{{Cocconi} and {Morrison}(1959)}]{SETI}
{Cocconi} G {Morrison} P (1959) {Searching for Interstellar Communications}.
  \nat 184:844--846

\bibitem[{{Corbet}(2003)}]{Corbet03}
{Corbet} RHD (2003) {Synchronized SETI-The Case for ``Opposition''}.
  Astrobiology 3:305--315

\bibitem[{{Cowan} and {Fujii}(2017)}]{Cowan17}
{Cowan} NB {Fujii} Y (2017) {Mapping Exoplanets}. ArXiv e-prints

\bibitem[{{Cowley} et~al.(2000){Cowley}, {Ryabchikova}, {Kupka}, {Bord},
  {Mathys}, and {Bidelman}}]{Cowley00}
{Cowley} CR, {Ryabchikova} T, {Kupka} F et~al. (2000) {Abundances in
  Przybylski's star}. \mnras 317:299--309

\bibitem[{{de Wit} et~al.(2012){de Wit}, {Gillon}, {Demory}, and
  {Seager}}]{deWit12}
{de Wit} J, {Gillon} M, {Demory} BO {Seager} S (2012) {Towards consistent
  mapping of distant worlds: secondary-eclipse scanning of the exoplanet HD
  189733b}. \aap 548:A128

\bibitem[{{Dixon}(1973)}]{Dixon73}
{Dixon} RS (1973) {A Search Strategy for Finding Extraterrestrial Radio
  Beacons}. \icarus 20:187--199

\bibitem[{{Dressing} and {Charbonneau}(2015)}]{Dressing15}
{Dressing} CD {Charbonneau} D (2015) {The Occurrence of Potentially Habitable
  Planets Orbiting M Dwarfs Estimated from the Full Kepler Dataset and an
  Empirical Measurement of the Detection Sensitivity}. \apj 807:45

\bibitem[{{Dyson}(1960)}]{Dyson60}
{Dyson} FJ (1960) {Search for Artificial Stellar Sources of Infrared
  Radiation}. Science 131:1667--1668

\bibitem[{{Filippova} and {Strelnitskij}(1988)}]{Filippova88}
{Filippova} LN {Strelnitskij} VS (1988) {Ecliptic as an Attractor for SETI}.
  Astronomicheskij Tsirkulyar 1531:31

\bibitem[{{Filippova} et~al.(1991){Filippova}, {Kardashev}, {Likhachev}, and
  {Strelnitskij}}]{Filippova91}
{Filippova} LN, {Kardashev} NS, {Likhachev} SF {Strelnitskij} VS (1991) {On the
  Strategy of SETI}. In: {Heidmann} J {Klein} MJ (eds) Bioastronomy: The Search
  for Extraterrestrial Life---The Exploration Broadens, Lecture Notes in
  Physics, Berlin Springer Verlag, vol 390, pp 254--258,
  \doi{10.1007/3-540-54752-5_225}

\bibitem[{{Forgan}(2013)}]{Forgan13}
{Forgan} DH (2013) {On the Possibility of Detecting Class A Stellar Engines
  using Exoplanet Transit Curves}. Journal of the British Interplanetary
  Society 66:144--154

\bibitem[{{Freeman} and {Lampton}(1975)}]{Freeman75}
{Freeman} J {Lampton} M (1975) {Interstellar Archaeology and the Prevalence of
  Intelligence}. \icarus 25:368--369

\bibitem[{{Guillochon} and {Loeb}(2015)}]{Guillochon15}
{Guillochon} J {Loeb} A (2015) {SETI via Leakage from Light Sails in
  Exoplanetary Systems}. \apjl 811:L20

\bibitem[{{Harp} et~al.(2016){Harp}, {Richards}, {Tarter}, {Dreher}, {Jordan},
  {Shostak}, {Smolek}, {Kilsdonk}, {Wilcox}, {Wimberly}, {Ross}, {Barott},
  {Ackermann}, and {Blair}}]{Harp16_exoplanets}
{Harp} GR, {Richards} J, {Tarter} JC et~al. (2016) {SETI Observations of
  Exoplanets with the Allen Telescope Array}. \aj 152:181

\bibitem[{{Henry} et~al.(1995){Henry}, {Soderblom}, {Baliunas}, {Davis},
  {Donahue}, {Latham}, {Stefanik}, {Torres}, {Duquennoy}, {Mayor}, {Andersen},
  {Nordstrom}, and {Olsen}}]{Henry95}
{Henry} T, {Soderblom} D, {Baliunas} S et~al. (1995) {The Current State of
  Target Selection for NASA's High Resolution Microwave Survey}. In: {Shostak}
  GS (ed) Progress in the Search for Extraterrestrial Life., Astronomical
  Society of the Pacific Conference Series, vol~74, p 207

\bibitem[{{Imara} and {Di Stefano}(2017)}]{Imara17}
{Imara} N {Di Stefano} R (2017) {Searching for Exoplanets Around X-Ray Binaries
  with Accreting White Dwarfs, Neutron Stars, and Black Holes}. arXiv:170305762

\bibitem[{{Isaacson} et~al.(2017){Isaacson}, {Siemion}, {Marcy}, {Lebofsky},
  {Price}, {MacMahon}, {Croft}, {DeBoer}, {Hickish}, {Werthimer}, {Sheikh},
  {Hellbourg}, and {Enriquez}}]{Isaacson17}
{Isaacson} H, {Siemion} APV, {Marcy} GW et~al. (2017) {The Breakthrough Listen
  Search for Intelligent Life: Target Selection of Nearby Stars and Galaxies}.
  ArXiv e-prints

\bibitem[{{Kasting} et~al.(1993){Kasting}, {Whitmire}, and
  {Reynolds}}]{Kasting93}
{Kasting} JF, {Whitmire} DP {Reynolds} RT (1993) {Habitable Zones around Main
  Sequence Stars}. Icarus 101:108--128

\bibitem[{{Kawahara} and {Fujii}(2010)}]{Kawahara10}
{Kawahara} H {Fujii} Y (2010) {Global Mapping of Earth-like Exoplanets From
  Scattered Light Curves}. \apj 720:1333--1350

\bibitem[{{Kessler} and {Cour-Palais}(1978)}]{Kessler78}
{Kessler} DJ {Cour-Palais} BG (1978) {Collision frequency of artificial
  satellites: The creation of a debris belt}. \jgr 83:2637--2646

\bibitem[{{Kipping} and {Teachey}(2016)}]{Kipping16}
{Kipping} DM {Teachey} A (2016) {A cloaking device for transiting planets}.
  \mnras 459:1233--1241

\bibitem[{{Knutson} et~al.(2007){Knutson}, {Charbonneau}, {Allen}, {Fortney},
  {Agol}, {Cowan}, {Showman}, {Cooper}, and {Megeath}}]{Knutson07}
{Knutson} HA, {Charbonneau} D, {Allen} LE et~al. (2007) {A map of the day-night
  contrast of the extrasolar planet HD 189733b}. \nat 447:183--186

\bibitem[{{Korpela} et~al.(2015){Korpela}, {Sallmen}, and {Leystra
  Greene}}]{Korpela15}
{Korpela} EJ, {Sallmen} SM {Leystra Greene} D (2015) {Modeling Indications of
  Technology in Planetary Transit Light Curves--Dark-side Illumination}. \apj
  809:139

\bibitem[{{Kreidberg} and {Loeb}(2016)}]{Kreidberg16}
{Kreidberg} L {Loeb} A (2016) {Prospects for Characterizing the Atmosphere of
  Proxima Centauri b}. \apjl 832:L12

\bibitem[{Kuhn and Berdyugina(2015)}]{kuhn2015global}
Kuhn JR Berdyugina SV (2015) Global warming as a detectable thermodynamic
  marker of earth-like extrasolar civilizations: the case for a telescope like
  colossus. International journal of astrobiology 14(3):401

\bibitem[{{Lin} et~al.(2014){Lin}, {Gonzalez Abad}, and {Loeb}}]{Lin14}
{Lin} HW, {Gonzalez Abad} G {Loeb} A (2014) {Detecting Industrial Pollution in
  the Atmospheres of Earth-like Exoplanets}. \apjl 792:L7

\bibitem[{{Lingam} and {Loeb}(2017)}]{Lingam17}
{Lingam} M {Loeb} A (2017) {Natural and artificial spectral edges in
  exoplanets}. \mnras 470:L82

\bibitem[{{Loeb} and {Turner}(2012)}]{Loeb11}
{Loeb} A {Turner} EL (2012) {Detection Technique for Artificially Illuminated
  Objects in the Outer Solar System and Beyond}. Astrobiology 12:290--294

\bibitem[{{Maire} et~al.(2016){Maire}, {Wright}, {Dorval}, {Drake}, {Duenas},
  {Isaacson}, {Marcy}, {Siemion}, {Stone}, {Tallis}, {Treffers}, and
  {Werthimer}}]{Maire16}
{Maire} J, {Wright} SA, {Dorval} P et~al. (2016) {A near-infrared SETI
  experiment: commissioning, data analysis, and performance results}. In:
  Society of Photo-Optical Instrumentation Engineers (SPIE) Conference Series,
  \procspie, vol 9908, p 990810, \doi{10.1117/12.2232861}

\bibitem[{{Majeau} et~al.(2012){Majeau}, {Agol}, and {Cowan}}]{Majeau12}
{Majeau} C, {Agol} E {Cowan} NB (2012) {A Two-dimensional Infrared Map of the
  Extrasolar Planet HD 189733b}. \apjl 747:L20

\bibitem[{{Makovetskii}(1977)}]{Makovetskii77}
{Makovetskii} PV (1977) {Nova Cygni - A synchrosignal for extraterrestrial
  civilizations}. \azh 54:449--451

\bibitem[{{Makovetskii}(1980)}]{Makovetskii80}
{Makovetskii} PV (1980) {Mutual strategy of search for CETI call signals}.
  \icarus 41:178--192

\bibitem[{{Oliver}(1979)}]{Oliver79}
{Oliver} BM (1979) {Rationale for the water hole}. Acta Astronautica 6:71--79

\bibitem[{{Osmanov}(2016)}]{Osmanov16}
{Osmanov} Z (2016) {On the search for artificial Dyson-like structures around
  pulsars}. International Journal of Astrobiology 15:127--132

\bibitem[{{Pace} and {Walker}(1975)}]{Pace75}
{Pace} GW {Walker} JCG (1975) {Time markers in interstellar communication}.
  \nat 254:400

\bibitem[{{Panov} et~al.(2014){Panov}, {Filippova}, and {Rudnitskii}}]{Panov14}
{Panov} A, {Filippova} L {Rudnitskii} G (2014) {Limited list for the objects -
  candidates for SETI-monitoring with large telescopes}. In: 40th COSPAR
  Scientific Assembly, COSPAR Meeting, vol~40

\bibitem[{{Petigura} et~al.(2013){Petigura}, {Howard}, and
  {Marcy}}]{Petigura13}
{Petigura} EA, {Howard} AW {Marcy} GW (2013) {Prevalence of Earth-size planets
  orbiting Sun-like stars}. Proceedings of the National Academy of Science
  110:19,273--19,278

\bibitem[{{Przybylski}(1961)}]{Przybylski61}
{Przybylski} A (1961) {HD 101065-a G0 Star with High Metal Content}. \nat
  189:739

\bibitem[{Schelling(1960)}]{schelling1960strategy}
Schelling T (1960) The strategy of conflict. Galaxy book, Harvard University
  Press, \urlprefix\url{https://books.google.com/books?id=Ctl-AAAAMAAJ}

\bibitem[{{Schneider} et~al.(2010){Schneider}, {L{\'e}ger}, {Fridlund},
  {White}, {Eiroa}, {Henning}, {Herbst}, {Lammer}, {Liseau}, {Paresce},
  {Penny}, {Quirrenbach}, {R{\"o}ttgering}, {Selsis}, {Beichman}, {Danchi},
  {Kaltenegger}, {Lunine}, {Stam}, and {Tinetti}}]{Schneider10}
{Schneider} J, {L{\'e}ger} A, {Fridlund} M et~al. (2010) {The Far Future of
  Exoplanet Direct Characterization}. Astrobiology 10:121--126

\bibitem[{Shklovski{\u\i} and Sagan(1966)}]{shklovskii1966}
Shklovski{\u\i} I Sagan C (1966) Intelligent life in the universe: Vselennaja
  zizn'razum. Delta-books, Holden-Day,
  \urlprefix\url{https://books.google.com/books?id=o4cRAQAAIAAJ}

\bibitem[{{Shostak}(2004)}]{Shostak04}
{Shostak} S (2004) {A Scheme for Targeting Optical SETI Observations}. In:
  {Norris} R {Stootman} F (eds) Bioastronomy 2002: Life Among the Stars, IAU
  Symposium, vol 213, p 409

\bibitem[{{Siemion} et~al.(2013){Siemion}, {Demorest}, {Korpela}, {Maddalena},
  {Werthimer}, {Cobb}, {Howard}, {Langston}, {Lebofsky}, {Marcy}, and
  {Tarter}}]{Siemion2013}
{Siemion} APV, {Demorest} P, {Korpela} E et~al. (2013) {A 1.1-1.9 GHz SETI
  Survey of the Kepler Field. I. A Search for Narrow-band Emission from Select
  Targets}. \apj 767:94

\bibitem[{{Singer}(1982)}]{Singer82}
{Singer} CE (1982) {When to Look Where}. Cosmic Search 4:22

\bibitem[{{Stevens} et~al.(2016){Stevens}, {Forgan}, and {James}}]{Stevens16}
{Stevens} A, {Forgan} D {James} JO (2016) {Observational signatures of
  self-destructive civilizations}. International Journal of Astrobiology
  15:333--344

\bibitem[{{Tarter}(2001)}]{Tarter01}
{Tarter} J (2001) {The Search for Extraterrestrial Intelligence (SETI)}. \araa
  39:511--548

\bibitem[{{Traub}(2012)}]{Traub12}
{Traub} WA (2012) {Terrestrial, Habitable-zone Exoplanet Frequency from
  Kepler}. \apj 745:20

\bibitem[{{Turnbull} and {Tarter}(2003{\natexlab{a}})}]{Turnbull03a}
{Turnbull} MC {Tarter} JC (2003{\natexlab{a}}) {Target Selection for SETI. I. A
  Catalog of Nearby Habitable Stellar Systems}. \apjs 145:181--198

\bibitem[{{Turnbull} and {Tarter}(2003{\natexlab{b}})}]{Turnbull03b}
{Turnbull} MC {Tarter} JC (2003{\natexlab{b}}) {Target Selection for SETI. II.
  Tycho-2 Dwarfs, Old Open Clusters, and the Nearest 100 Stars}. \apjs
  149:423--436

\bibitem[{{Whitmire} and {Wright}(1980)}]{Whitmire80}
{Whitmire} DP {Wright} DP (1980) {Nuclear waste spectrum as evidence of
  technological extraterrestrial civilizations}. \icarus 42:149--156

\bibitem[{{Wolszczan} and {Frail}(1992)}]{Wolszczan92}
{Wolszczan} A {Frail} DA (1992) {A planetary system around the millisecond
  pulsar PSR1257 + 12}. \nat 355:145--147

\bibitem[{{Wright} et~al.(2014{\natexlab{a}}){Wright}, {Griffith},
  {Sigurdsson}, {Povich}, and {Mullan}}]{GHAT2}
{Wright} JT, {Griffith} RL, {Sigurdsson} S, {Povich} MS {Mullan} B
  (2014{\natexlab{a}}) {The {\^G} Infrared Search for Extraterrestrial
  Civilizations with Large Energy Supplies. II. Framework, Strategy, and First
  Result}. \apj 792:27

\bibitem[{{Wright} et~al.(2014{\natexlab{b}}){Wright}, {Mullan}, {Sigurdsson},
  and {Povich}}]{GHAT1}
{Wright} JT, {Mullan} B, {Sigurdsson} S {Povich} MS (2014{\natexlab{b}}) {The
  {\^G} Infrared Search for Extraterrestrial Civilizations with Large Energy
  Supplies. I. Background and Justification}. \apj 792:26

\bibitem[{{Wright} et~al.(2016){Wright}, {Cartier}, {Zhao}, {Jontof-Hutter},
  and {Ford}}]{GHAT4}
{Wright} JT, {Cartier} KMS, {Zhao} M, {Jontof-Hutter} D {Ford} EB (2016) {The
  Search for Extraterrestrial Civilizations with Large Energy Supplies. IV. The
  Signatures and Information Content of Transiting Megastructures}. \apj 816:17

\end{thebibliography}

\end{document}